\documentstyle[preprint,eqsecnum,aps,floats]{revtex}
\begin{document}

%
\begin{titlepage}
\title{BI action for gravitational and electroweak  fields}
\author{Dmitriy Palatnik\thanks{{ }6400 N. Sheridan Rd. Apt. 715 Chicago, IL 60626 ; 
mailto: palatnik@ripco.com}
}

\maketitle

\tighten
\begin{abstract}

This note suggests a generalization of the Born--Infeld action (1932)
on case of electroweak and gravitational fields in four-dimensional spacetime.
The action is constructed from Dirac matrices, $\gamma_a$, and dimensionless covariant
derivatives, $\pi_{a} = - i\ell \nabla_{a}$, where
$\ell$ is of order of magnitude of Planck's length.
By a postulate, the action possesses additional symmetry with respect to global transformations
of the Lorentz group imposed on pairs ($\gamma_{a}$, $\pi_{a}$).
It's shown, that parameter of the Lorentz group is associated with
a constant value of the electroweak potential at spatial infinity.
It follows, that in  $\ell^6$ approximation, action for gravitational
field  coincides with  Einstein-Hilbert (EH) and Gauss-Bonnet (GB).  

\end{abstract}

\vspace{20pt}
\center{PACS numbers: 11.10.Lm; 12.10.-g; 12.60.-i}

\end{titlepage}

\tighten

\section{Introduction}

Development in string theory has caused resurrection
of  the Born--Infeld (BI) model \cite{BI,GWG,k}, initially formulated for
electromagnetic fields. It was shown \cite{ft,bs,mr,ts}, that
for open strings the low-energy effective actions
coincide with the BI one. 
Couplings of the electromagnetic BI  to the EH action 
were studied in, e.g., \cite{gh,wsj,abg1,cmm}.  
Nonlinear in curvature tensor gravitational actions 
were considered in, e.g.,  \cite{ffp,DG,w,n,9701017}.

In this paper one constructs `from scratch' (i.e., with no reference to any specific string
model) a BI action for gravitational and electroweak fields. As the leading principle
one accepts additional symmetry of the action under global Lorentz transformations
imposed on pairs $(\gamma_a, \pi_a)$ where $\gamma_a$ are the Dirac matrices 
and $\pi_a$ are dimensionless covariant derivative operators. 
This symmetry narrows set of invariants, eligible for entering into the action.

The key feature of the suggested action
is its nonlinearity  in curvature tensor, $\rho_{ab}$; the latter
substitutes, roughly speaking, dimensionless electromagnetic tensor, $F_{ab}$, 
in BI invariant, $\int d\Omega \sqrt{-\det(\eta_{ab} + F_{ab})}$.
Due to that nonlinearity one obtains both linear and quadratic in $\rho_{ab}$
terms in the action's expansion, corresponding, respectively, to the 
EH and GB terms for gravity and to the standard one for electroweak fields. 

One is able  to conceal all fields' potentials in connections,
associated with $\pi_a$. As a result, all interactions
enter the action on equal footing. 

To begin with, one specifies basic elements of the action,
compatible with quantum field theory in four-dimensional spacetime; namely,   

($\alpha$) Dirac $n \times n$ matrices, $\gamma_{a}$, submitted to relations
\begin{eqnarray}\label{gdef}
\gamma_{(a}\gamma_{b)} & = & g_{ab}\,\hat{1}_n\,,
\end{eqnarray}
where $g_{ab}$ is the metric tensor, and $\hat{1}_n$ is unit $n \times n$ matrix.  
It is assumed below, if not specified differently,  that tensor indices, $a, b, c,$... are 
lowered and raised with
$g_{ab}$ and $g^{ab}$, respectively, where $g^{ab}g_{cb} = \delta^a_c$.

($\beta$) Dimensionless operators
\begin{equation}\label{pi}
\pi_a \,=\, -i\ell\nabla_a\,,
\end{equation}
where $\nabla_a$ is a covariant derivative and $\ell$ is characteristic length.
Action of (\ref{pi}) on spinors, scalars, and Dirac matrices
is specified in the following text. 
For spinors (scalars) e.g., one obtains,
$\pi_a\Psi   =   -i\ell\left({\partial}_a\Psi - \Gamma_{a}\Psi\right)$;
here $\Psi$  represents spinor (scalar), and $\Gamma_a$  
are respective connection matrices. Introducing $\Psi_a = \gamma_a\Psi$, one obtains,
$\pi_b\Psi_a   =   -i\ell\left({\partial}_b\Psi_a - \Gamma_{ab}^c\Psi_c - \Gamma_{b}\Psi_a\right)$.
Here $\Gamma_{ab}^c = \Gamma_{(ab)}^c$ is connection symbol.\footnote{{ }%
The connection symbol, $\Gamma_{ab}^c$, coincides with the Christoffel one in zero
approximation in $\ell^2$.  One's intent is to use the Palatini method in order to find $\Gamma$'s 
and $\gamma$'s.}

($\gamma$) Curvature tensor,
\begin{eqnarray}\label{curvature}
{\rho}_{ab} & = & 2\ell^{-2}\,\pi_{[a}\pi_{b]} =  2\left(\partial_{[a}\Gamma_{b]} -
\Gamma_{[a}\Gamma_{b]}\right)\,,
\end{eqnarray}
where $\Gamma_a$ are connections, given in spinorial (scalar) representation.

\section{Symmetry and objects}

By a postulate, the action is based on global Lorentz-invariant combinations of
$\gamma_a$ and $\pi_a$. Namely,
action is invariant under substitutions $\gamma_a \rightarrow \gamma_a'$,
 $\pi_a \rightarrow \pi_a'$, where
\begin{eqnarray}\label{symm}
      \gamma_a' & = &
\cosh\theta\,\gamma_{a} + \sinh\theta\,\pi_a\,,\\
\label{symm1} \pi_a' & = &
\sinh\theta\,\gamma_{a} + \cosh\theta\,\pi_a\,,
\end{eqnarray}
and $\theta$ doesn't depend on coordinates.\footnote{{ }%
It is shown below that $\theta$ is associated with the value of electroweak
potential at spatial infinity.}

1. Define  first group of operators, $p_{ab} = p_{[ab]}$, $p$, and $p^{ab} = p^{[ab]}$,
\begin{eqnarray}\label{phiop}
p_{ab} & = & \gamma_{[a}\gamma_{b]} - \pi_{[a}\pi_{b]}\,;\\
\label{phi}p & = & {1\over
{4!}}\,e^{abcd}\,e^{efgh}\,{1\over3}\left[ n^{-1}\,{tr}(p_{ae}\,
p_{bf}\,p_{cg}\,p_{dh}) - 2 n^{-2}\, tr(p_{ae}\,
p_{bf})\,tr(p_{cg}\,p_{dh})\right]\,;\\
\label{contra}{p}^{dh} & = &
{1\over{3!\,p}}\,e^{abcd}e^{efgh}\,{1\over3}\left[p_{ae}\,
p_{bf}\,p_{cg} - 2 n^{-1}\, tr(p_{ae}\,
p_{bf})\,p_{cg}\right]   \,; { }
\end{eqnarray}
here and below $e^{abcd} = e^{[abcd]}$, and $e_{abcd} = e_{[abcd]}$ are the absolute 
antisymmetric symbols, $e^{0123} = 1$, and $e_{0123} = -1$.

2. Define second group of operators, $q_{ab} = q_{(ab)}$, $q$, 
and $q^{ab} = q^{(ab)}$,
\begin{eqnarray}\label{chi1}
q_{ab} & = & \gamma_{(a}\gamma_{b)} - \pi_{(a}\pi_{b)}\,;\\
\label{detchi}q & = & {1\over
{4!}}\,e^{abcd}\,e^{efgh}\,(q_{ae}\,
q_{bf}\,q_{cg}\,q_{dh})\,;\\
\label{chi1inv}
q^{dh} &=& {1\over{3!}}\,q^{-1}e^{abcd}\,e^{efgh}\,(q_{ae}\,
q_{bf}\,q_{cg})\,.
\end{eqnarray}
{}From (\ref{chi1}) and (\ref{gdef}) follows, $q_{ab} = g_{ab} - \pi_{(a}\pi_{b)}$.

3. The third group of operators consists of $S_{ab} = S_{[ab]}$,
\begin{eqnarray}\label{sigma}
S_{ab} & = & \gamma_{[a}\pi_{b]} - \pi_{[a}\gamma_{b]}\,.
\end{eqnarray}

Scalar densities, reducing to $\sqrt{-g}$ in the limit
$\ell\rightarrow0$, are $\sqrt{-p}$ and $\sqrt{-q}$.\footnote{{ }%
One should understand  $q^{-1}$ and $\sqrt{-q}$ as 
 respective expansions in powers of $\ell^2$.}
Objects, defined in (\ref{phiop}) -- (\ref{sigma}),
are invariant with respect to `$\theta$-rotations,' (\ref{symm}) and (\ref{symm1}).

4. One may prove a useful formula, following from (\ref{gdef}),
\begin{eqnarray}\label{myform}
e^{abcd}\,e^{efgh}\,\gamma_{[a}\gamma_{e]}\,\gamma_{[b}\gamma_{f]}\,
\gamma_{[c}\gamma_{g]} & = & 10\,g\,\gamma^{[h}\gamma^{d]}\,.
\end{eqnarray}

\section{Action for the fields}

1. Take a system including electron, neutrino, electroweak
and gravitational fields, together with scalar fields, quadruplets $U$ and $V$.
Introduce two spinorial bases. First, $L$-basis, is represented by
octets ($n  = 8$),
$L_{el}   =  \lbrack \nu_L\,, e_L\rbrack^T$,
$\overline{L}_{el} = \lbrack \overline{\nu}_L\,,
\overline{e}_L\rbrack$.
Here $e_L$ and $\nu_L$ are 4-spinors for the left-handed electron and
neutrino, respectively. 
The second, $R$-basis, is represented by quadruplets, ($n  = 4$),
$e_R$ and $\overline{e}_R$,
corresponding to a right-handed electron.

2. One may specify action of $\pi_a$ on introduced fields.
\begin{eqnarray}\label{cderiv1}
\pi_aL_{el} & = & -i\ell\left(\partial_a +
{i\over2}g^{\prime}\,\sigma_k\otimes\hat1_4\,W_a^k
+ {i\over 2}g^{\prime\prime}\,{Y}_L\otimes\hat1_4\,B_a -
\hat1_2\otimes{\overline\Gamma}_a\right)L_{el}\,,\\
\label{cderiv2}\pi_ae_R & = & -i\ell\left(\partial_a +
{i\over 2}g^{\prime\prime}\,{Y}_R\,B_a -
{\overline\Gamma}_a\right)e_R\,,\\
\label{cderiv3}\pi_aU & = & -i\ell\left(\partial_a +
{i\over2}g^{\prime}\,\hat{1}_2\otimes \sigma_k\,W_a^k
+ {i\over 2}g^{\prime\prime}\,{Y}_W\,B_a
\right)U\,,\\
\label{cderiv4}\pi_aV& = & -i\ell\,\partial_a V\,.
\end{eqnarray}
Here $B_a$ and $W_a^k$, $k = 1,\,2,\,3$, are potentials of
electroweak fields; ${Y}_L = -\hat{1}_2$,
${Y}_R = -2\times\hat1_4$, and $Y_W = \hat{1}_4$ are hypercharge
operators;
$\sigma_k$, $k = 1,\,2,\,3$, are standard $2\times2$
Pauli matrices; and $g^{\prime}$ and $g^{\prime\prime}$ are interaction
constants.
Connections ${\overline\Gamma}_a$ are matrices $4\times 4$,
pertaining to gravitational fields only.
The Dirac matrices,
$\gamma_{a}$, have the following structure. In $L$-basis,
$\gamma_{a}  = $ diag$\,\left({\overline\gamma}_a\,,
{\overline\gamma}_a\,\right)
       =  \hat1_2\otimes{\overline\gamma}_a$,
where ${\overline\gamma}_a$ are Dirac matrices $4\times4$.
In $R$-basis, as well as in scalars' $U$ and $V$ bases, $\gamma_a =
{\overline\gamma}_a$.

3. To be more specific about the gravitational sector, one may introduce four
standard Dirac matrices $4\times4$, $\Delta_A$, $A = 0, 1, 2, 3$:
\begin{eqnarray}
\Delta_0 & = & \left(\begin{array}{cc}
          0 & \hat1_2 \\
          \hat1_2 & 0 \end{array}\right)\,,\,\,\,
\Delta_k  =  \left(\begin{array}{cc}
          0 & \sigma_k\\
          - \sigma_k & 0 \end{array}\right)\,,\;  k = 1, 2, 3.\nonumber
\end{eqnarray}
Then, the following decompositions take place:
\begin{eqnarray}\label{tetrad}
{\overline\gamma}_a & = & (e_a)^A\Delta_A\,,\\
\label{f_fields}{\overline\Gamma}_a & = & {1\over4}\,
(f_a)^{AB}\Delta_{[A}\Delta_{B]}\,,
\end{eqnarray}
where $(e_a)^A$ is a tetrad, and $(f_a)^{AB} = (f_a)^{[AB]}$
are six vector fields, which are associated with  Ricci
rotation coefficients \cite{fok} in standard theory.\footnote{{ }%
Namely,
$(f_a)^{AB} = (e_b)^B{\nabla}_a\,(e^b)^{A}$, where
${\nabla}_a$ is a covariant derivative, associated with metrics
$g_{ab} = (e_a)^A(e_b)_{A}$.}
 It is assumed below, that tetrad indices, 
$A, B, C,$ ... are lowered and raised with `Minkowski metrics,'
$\eta_{AB}$ and $\eta^{AB}$, respectively.

4. Expanding (\ref{phi})  in powers of $\ell^2$,
and using (\ref{myform}), one obtains for two spinorial bases the following
decompositions. For the $L$-basis,
\begin{eqnarray}\label{exp_phi_L}
\sqrt{-p_L} & = & \sqrt{-g} + {{\ell^2}\over{4!2}}\sqrt{-g}\,
tr( \overline{\gamma}^a\overline{\gamma}^b
\overline{\rho}_{ab}) 
+ {{\ell^4}\over{(4!)^2}}\,\sqrt{-g}\left(
g^{\prime2}W_k^{ab}W_{ab}^k + g^{\prime\prime2}B^{ab}B_{ab}\right)
+ O(\ell^4)\,.
\end{eqnarray}
For the $R$-basis,
\begin{eqnarray}\label{exp_phi_R}
\sqrt{-p_R} & = & \sqrt{-g} + {{\ell^2}\over{4!2}}\sqrt{-g}\,
tr( \overline{\gamma}^a\overline{\gamma}^b
\overline{\rho}_{ab})
+ {{\ell^4}\over{4!3!}}\,\sqrt{-g}\,
g^{\prime\prime2}B^{ab}B_{ab} + O(\ell^4)\,,
\end{eqnarray}
where $\overline{\rho}_{ab}$ is constructed via $\overline{\Gamma}_a$
as in (\ref{curvature}).
One uses notations $W_{ab}^k = \partial_aW_b^k -  \partial_bW_a^k -
g^{\prime}\epsilon_{lm}^kW_a^lW_b^m$
      and $B_{ab} = \partial_aB_b - \partial_bB_a$ for
electroweak gauge fields.\footnote{{ }Indices k, l, m ... are raised
and lowered with $\delta_{km} = \delta^{km} =$ diag$(1, 1, 1)$.}
As shown below, the second term in the right hand side of (\ref{exp_phi_L})
and  (\ref{exp_phi_R}) corresponds to EH term.
As it follows from (\ref{phiop}), (\ref{phi}), and (\ref{cderiv4}),
\begin{equation}\label{phi_V}
\sqrt{-p_{V}}\, =\, \sqrt{-g}\,,
\end{equation}
where $p_{V}$ denotes $p$ (\ref{phi}), given in
$V$-representation.

5.  From structure of (\ref{cderiv3}) follows that under respective matrix
transformations components $U' = [u_1,\,u_2]$ of the scalar
$U = [u_1,\,u_2,\, u_3,\,u_4]^T$
do not mix with components $U'' = [u_3,\,u_4]$ and vice versa, which
means that one
may consider two independent doublets, $[U',\,U'']$, combined in $U$.
For implementation of the symmetry breaking mechanism in order to generate 
masses for vector bosons and fermions, one may use scalar field,  $U$.
The respective potential energy,
\begin{equation}\label{penergy}
W(\overline{U},U,\overline{V},V) \,=\,
\sigma_U\, \overline{U}U + \kappa_U\, (\overline{U}U)^2 +
\sigma_V\,\overline{V}V\,,
\end{equation}
where $\sigma_U$, $\sigma_V$, and $\kappa_U$ are constants.
In the minimum of potential energy (\ref{penergy}) scalar fields may be taken with
the following constant values: $U' = [0,\,a]^T$, $U'' = [0,\,b]^T$, and
$\overline{U}' = [0,\,a]$, $\overline{U}'' = [0,\,b]$;
one obtains, then,
\begin{eqnarray}\label{chiex}
 \overline{U}\sqrt{-q}U & = & v\sqrt{-g}  - \sqrt{-g}\,{{v\ell^2}\over8}
g'^2\left(W_a^1W^{1a} + W_a^2W^{2a}\right)\nonumber\\
& & - \sqrt{-g}\,{{v\ell^2}\over8}\left(g'W_a^3 - g''B_a\right)
\left(g'W^{3a} - g''B^a\right) + O(\ell^2)\,.
\end{eqnarray}
Here $v = a^2 + b^2$.
Similarly, one may write an expansion
\begin{eqnarray}\label{chiexv}
\overline{V}\sqrt{-q}V & = & \sqrt{-g}\left( \overline{V}V
+ {{\ell^2}\over2}\,
\overline{V}\partial_a\partial^a V \right) + O(\ell^4)\,.
\end{eqnarray}
Note, that scalar field's $V$ sole purpose is to make invariant (\ref{phi_V}) 
possible; in turn, one needs $\sqrt{-p_V}$ in order to adjust the `cosmological term,' so
that it would have a reasonable value.
The field $V$ doesn't interact with other fields. On the other hand, it contributes stress-energy
to the Einstein equations; thus one may think of it as of the `dark matter.'

6. Comparing (\ref{exp_phi_L}) -- (\ref{chiexv}), one defines action for the fields,
\begin{eqnarray}\label{S_f}
S_f & = & \int d\Omega \nonumber\\
& & \times
\left[\lambda_L\sqrt{-p_L} + \lambda_R\sqrt{-p_R} -
  \Upsilon\,\sqrt{-p_{V}} +
\lambda_{U}\overline{U}\sqrt{-q}U  +
\lambda_{V}\overline{V}\sqrt{-q}V\right]\,,
\end{eqnarray}
where one introduces  constants,
$\lambda_L, \lambda_R, \lambda_{U}, \lambda_{V}$,
and scalar $\Upsilon$, containing the potential (\ref{penergy}),
necessary for symmetry-breaking mechanism:
\begin{eqnarray}\label{potential}
\Upsilon & = &\lambda_L + \lambda_R  + W\,.
\end{eqnarray}
The action (\ref{S_f}) together with the action for spinors
(\ref{S_psi}) should be varied with respect to fields
$(e_a)^A,\,(f_a)^{AB},\,W_a^k,\,B_a,\,\overline{U},\, U,\,
\overline{V},\, V ,\,
\overline{L}_{el}$, $L_{el}$, $\overline{e}_R$, and $e_R$.

Comparing the expansion of (\ref{S_f}) in powers of $\ell^2$ with the standard action,
one obtains the following expressions for constants:
\begin{eqnarray}
\label{ell}
\ell^2 & = & 6(1 + 2\sin^2\theta_W)\,\alpha^{-1}\ell_p^2\,;\\
\label{lambda_L}
\lambda_L & = & - {6\over{\pi}}\,{{\sin^2\theta_W}\over{1 + 2\sin^2\theta_W}}\,{{c^3}\over{k\ell^2}}\,;\\
\label{lambda_R}
\lambda_R & = & - {3\over{2\pi}}\,{{1 - 2\sin^2\theta_W}\over{1 + 2\sin^2\theta_W}}
\,{{c^3}\over{k\ell^2}}\,;\\
\label{lambda_U}
\lambda_{U} & = & - {1\over{8\pi c\ell^2}}\,;\\
\label{lambda_V}
\lambda_{V} & = & - {1\over{8\pi c\ell^2}}\,;\\
\label{mass}
a^2 + b^2 & = & 4\sin^2\theta_W\,{{m_W^2c^4}\over{e^2}}\,,
\end{eqnarray}
where $\alpha = e^2/\hbar c$ is the fine structure constant for an electron,
$\theta_W$ is Weinberg's mixing angle \cite{Halzen},
$\ell_p$ is Planck's length, and $m_W$ is vector bosons' rest mass.

\section{Action for the spinors}

Using invariants (\ref{contra}) and (\ref{sigma}), the action for fermions may be constructed as follows:
\begin{eqnarray}\label{S_psi}
S_{\Psi} & = & {\hbar\over{2\ell}} \int d\Omega \nonumber\\
& & \times \,\left[\sqrt{-p_L}\,
\overline{L}_{el}\,p^{ab}\,S_{ab} L_{el}
+ \sqrt{-p_R}\,\overline{e}_R\,p^{ab}\,S_{ab} e_R
      +  r\sqrt{-p_{V}}\,\left( \overline{L}_{el}e_RU' +
\overline{U}'\overline{e}_RL_{el}\right)\right]\,.
\end{eqnarray}
Expanding (\ref{S_psi}) in powers of $\ell$, one obtains,
\begin{eqnarray}\label{d_exp}
S_{\Psi} = \int d\Omega\sqrt{-g}\left[
\overline{L}_{el}\,\gamma^a{p}_a L_{el}
+ \overline{e}_R\,\overline{\gamma}^a{p}_a e_R
      + {{\hbar r a}\over{2\ell}}\left(
\overline{e}_Le_R + \overline{e}_Re_L\right)\right] + \cdots
\end{eqnarray}
The momentum operator, ${p}_a = -i\hbar (\partial_a - \Gamma_{a})$, is
given in respective spinorial bases.

\section{Gravitational sector}

Consider vacuum gravitational fields. The action reduces to the following,
\begin{equation}\label{gb7}
S_g \,=\, k \int d\Omega \left( \sqrt{-p} - \sqrt{-g}\right)\,.
\end{equation}
Here $\sqrt{-p} = (\lambda_L + \lambda_R)^{-1}( \lambda_L\sqrt{-p_L} + \lambda_R\sqrt{-p_R})$,
with electroweak fields $W_a^k = B_a = 0$. 
Using definition
$f^{ab} = \sqrt{p\over g}\,p^{ab}$, where $p$ and $p^{ab}$ are defined by
(\ref{phi}) and (\ref{contra}), one may obtain the field equations,\footnote{{ }%
In this section one uses notations $\gamma_a$ and $\Gamma_a$ instead of 
$\overline\gamma_a$ and $\overline\Gamma_a$, for clarity.
}
\begin{eqnarray}\label{sss1}
{1\over{\sqrt{-g}}}\,\partial_a\left[\sqrt{-g}\,f^{ba}\right] - [\Gamma_a, f^{ba}] &=& 0\,;\\
\label{sss2}
[f^{ab}, \gamma_a] & = & 2\gamma^b\,.
\end{eqnarray}

Define $L_{ab}^{cd} = L_{[ab]}^{[cd]}$ according to
\begin{equation}\label{Ldef}
L_{cd}^{ab} \,=\, {1\over2}(\delta^a_c \delta^b_d - \delta^a_d \delta^b_c) + {\ell^2\over 8}\,\tilde\rho^{ab}_{cd}\,,
\end{equation}
where 
\begin{equation}\label{gb1a}
\tilde\rho_{ab}^{cd} \,=\, \left[\partial_a(f_b)^{AB} - \partial_b(f_a)^{AB} +
(f_a)_C^{\;\;\,A}\,(f_b)^{CB} - (f_b)_C^{\;\;\,A}\,(f_a)^{CB}\right](e^c)_B(e^d)_A\,.
\end{equation}
Here one uses  definitions (\ref{tetrad}) and (\ref{f_fields}). As it is shown below,
$\tilde\rho_{abcd} = R_{abcd} + O(\ell^6)$, where $R_{abcd}$ is the Riemann tensor, 
and $\tilde\rho_{ab cd} = g_{ce}\,g_{df}\,\tilde\rho_{ab}^{ef}$. 
One obtains, 
\begin{equation}\label{pab}
p_{ab} \,= \,L_{ab}^{ef}\,\gamma_e\gamma_f\,.  
\end{equation}

One may show that in vacuum gravitational action in  $\ell^6$ approximation is sum 
of EH and GB terms,
\begin{eqnarray}\label{lagrangianstar}
\sqrt{-p} &=& \sqrt{-g}\,\lbrace 1 + {{\ell^2}\over{4!}} \,\tilde\rho 
  + {{3\ell^4}\over{2(4!)^2}}\,\left(\tilde\rho^2 - 4\tilde\rho^{ab}\tilde\rho_{ab}  + 
\tilde\rho^{abcd}\tilde\rho_{abcd}\right)\rbrace + O(\ell^8)\,.
\end{eqnarray}
Here $\tilde\rho_{ab} = g^{cd}\tilde\rho_{acbd}$, and $\tilde\rho = g^{ab}\tilde\rho_{ab}$.
To simplify calculations, one may choose a preferred coordinate frame, in which at an event
metric is Minkowski and Riemann tensor is brought to its canonical form. 
Assume that Riemann tensor at the event is of Petrov type I.
This means that up to terms of order $\ell^4$  tensor $\tilde\rho_{abcd}$  has only following non-zero 
components: $\tilde\rho^{01}_{01}, \tilde\rho^{02}_{02}, 
\tilde\rho^{03}_{03}, \tilde\rho^{12}_{12}, \tilde\rho^{13}_{13}$,  $\tilde\rho^{23}_{23}$, 
$\tilde\rho_{0123}, \tilde\rho_{0231}$, and $\tilde\rho_{0312}$.
Using (\ref{phi}) one may show,
\begin{eqnarray}\label{pdec}
p &=& {16\over9}\, g\,[ (L^{01}_{01}L^{23}_{23} + L^{02}_{02}L^{31}_{31} + L^{03}_{03}L^{12}_{12}
+ L^1L_1 +  L^2L_2 +  L^3L_3)^2 \nonumber\\
& & - (L_1x_1 + L_2x_2 + L_3x_3)(L^1x_1 + L^2x_2 + L^3x_3) \nonumber\\
 & & - 2(L^1L_1y_1^2 + L^2L_2y_2^2 + L^3L_3y_3^2)] + O(\ell^8)\,.
\end{eqnarray}
Here one uses notations, $L_1 = L_{0123}$, $L_2 = L_{0231}$,
$L_3 = L_{0312}$, $L^1 = L^{0123}$, $L^2 = L^{0231}$,
$L^3 = L^{0312}$; $x_1 = L^{01}_{01} + L^{23}_{23} - 1$, 
$x_2 = L^{02}_{02} + L^{31}_{31} - 1$, $x_3 = L^{03}_{03} + L^{12}_{12} - 1$;
$y_1 = L^{01}_{01} - L^{23}_{23}$, $y_2 = L^{02}_{02} - L^{31}_{31}$, $y_3 = L^{03}_{03} - L^{12}_{12}$.
One has also used indentity $L_1 + L_2 + L_3 = O(\ell^6)$, 
$L^1 + L^2 + L^3 = O(\ell^6)$.
By substituting definitions of $L^{cd}_{ab}$, one obtains from (\ref{pdec}), 
\begin{eqnarray}\label{sqrtp}
p &=& g\,[1 + {{2\ell^2}\over{4!}}(\tilde\rho^{01}_{01} +
\tilde\rho^{02}_{02} + \tilde\rho^{03}_{03} + \tilde\rho^{12}_{12} + \tilde\rho^{13}_{13} + \tilde\rho^{23}_{23})
\nonumber\\
& & + {{\ell^4}\over{4!2}}(\tilde\rho^{01}_{01}\tilde\rho^{23}_{23} + \tilde\rho^{02}_{02}\tilde\rho^{31}_{31}
+ \tilde\rho^{03}_{03}\tilde\rho^{12}_{12} +\tilde\rho^{01}_{23} \tilde\rho^{23}_{01}+ 
\tilde\rho^{02}_{31} \tilde\rho^{31}_{02} + \tilde\rho^{03}_{12} \tilde\rho^{12}_{03}
)]^2  + O(\ell^8)\,.
\end{eqnarray}
One may rewrite (\ref{sqrtp}) in covariant form. The first term
in parenthesis on the right hand side of (\ref{sqrtp}) is just ${1\over2}\tilde\rho$. 
The only quadratic in $\tilde\rho_{abcd}$ term, 
reducing to the second one in parenthesis on the right hand side of (\ref{sqrtp})
(under selected form of degeneracy of $\tilde\rho_{ab}^{cd}$) is exactly  GB invariant,
${1\over8}(\tilde\rho^2 - 4\tilde\rho^{ab}\tilde\rho_{ab}  + \tilde\rho^{abcd} \tilde\rho_{abcd})$. 
Thus, one obtains (\ref{lagrangianstar}), which should be valid in arbitrary coordinate system.

Similarly, one may show,
\begin{equation}\label{Pdec1}
f^{ab} \,=\, {1\over {12}}\,e^{abef}e_{cdgh}\,L_{ef}^{gh}\,\gamma^c\,\gamma^d + O(\ell^8)\,.
\end{equation}
{}From (\ref{phi}) and (\ref{contra}) follows $n^{-1}tr(p^{ab}p_{ab}) = 4$.  Using that result
and (\ref{pab}), (\ref{Pdec1}) together with definition of $f^{ab}$, one obtains yet another  
expression for $\sqrt{-p}$,
\begin{equation}\label{pinv}
\sqrt{-p} \,=\, - {1\over {4!}}\,\sqrt{-g}\,e^{abef}e_{cdgh}\,L_{ef}^{gh}\,L_{ab}^{cd} + O(\ell^8)\,.
\end{equation}
Substituting (\ref{Pdec1}) in (\ref{sss2}), one obtains,
\begin{equation}\label{Eeq}
\tilde\rho^a_b - {1\over2}\,\delta^a_b\,\tilde\rho \,=\, O(\ell^6)\,.
\end{equation}
Substituting (\ref{Pdec1}) in (\ref{sss1}), one obtains,
\begin{equation}\label{Eeq1}
(\gamma^c\gamma^d)_{;\,b}\,e^{abef}e_{cdgh}\,L^{gh}_{ef} +
\gamma^c\gamma^d\,e^{abef}e_{cdgh}\,L^{gh}_{ef;\,b} \,=\, O(\ell^8)\,.
\end{equation} 
Due to Bianchi identity, one obtains
\begin{equation}\label{bianchi}
e^{abef}\,L^{gh}_{ef;\,b} \, = \, O(\ell^8)\,.
\end{equation}
{}From (\ref{Eeq1}), (\ref{bianchi}) follows
\begin{eqnarray}\label{1up}
\gamma_{a\,;\,b} & = & O(\ell^8)\,.
\end{eqnarray}
For covariant derivative of ${\gamma}_a$ in (\ref{1up}) one obtains,
\begin{eqnarray}\label{D_gamma_def}
{\gamma}_{a\, ;\, b} & \equiv & {i\over {\ell}}\,[\pi_b, {\gamma}_a] 
 \,=\, \partial_b{\gamma}_{a}
- {\Gamma}_{ab}^c{\gamma}_{c} -
\left[ {\Gamma}_{b}\,, {\gamma}_{a} \right] \,.
\end{eqnarray}
Applying antisymmetrized product of covariant derivatives
to ${\gamma}_{b}$ and using (\ref{curvature}), (\ref{1up}), one obtains,\footnote{{ }%
Due to (\ref{Eeq}), the Christoffel symbol, $\Gamma_{ab}^c$, has corrections
of order $\ell^6$ to its vacuum values. Hence (\ref{riemann}) is calculated up to terms of order
$\ell^4$.
}
\begin{eqnarray}\label{riemann}
R_{\,\,\,bcd\,}^{\,a}{\gamma}_{a} & = & \lbrack{\gamma}_{b}\,, {\rho}_{cd}\rbrack + O(\ell^6)\,.
\end{eqnarray}
From (\ref{riemann}) one finds,
\begin{equation}\label{curvature_tensor}
\rho_{cd} \,=\, - {1\over 4}\,R_{abcd}\, \gamma^a\gamma^b + O(\ell^6)\,.
\end{equation}
Finally, from (\ref{phiop}), (\ref{Ldef}), (\ref{pab}), and (\ref{curvature_tensor})
follows, that $\tilde\rho_{abcd} = R_{abcd} + O(\ell^6)$.

{}From structure of (\ref{pdec}) it follows, that solutions of (\ref{sss1}) and (\ref{sss2}) with high symmetry,
namely, those possessing diagonal metric and `diagonal'  $\tilde\rho_{abcd} = R_{abcd}$ 
(i.e., one having $\tilde\rho_{abab}$ as the only non-zero components),
coincide with respective solutions of the Einstein's equations.

\section{Meaning of the $\theta$}

{}From the structure of the action for an electron (\ref{d_exp}), follow
expressions for correction to the energy, $\delta E$, due to the value
of the potentials of electroweak fields at spatial infinity.
Namely,
\begin{eqnarray}\label{delta_e1}
\delta E_L &=& {\hbar\over 2}\left(g'W_0^3(\infty) + g''B_0(\infty)\right)
\int dV\, \overline{e}_L\overline\gamma^0 e_L\,;\\
\label{delta_e2}
\delta E_R &=& \hbar g''B_0(\infty) \int dV
\,\overline{e}_R\overline\gamma^0 e_R\,.
\end{eqnarray}
One uses notations, $W_0^3(x^a) \rightarrow W_0^3(\infty)$,
as $|x^a| \rightarrow \infty$, etc.
{}From (\ref{symm}) and  (\ref{symm1}) follows,
\begin{eqnarray}\label{symm_int}
\int \overline{L}_{el}{\gamma'}^a  dS_a L_{el}& = &
\cosh\theta\,\int \overline{L}_{el}\gamma^a  dS_a L_{el}
+ \sinh\theta\, \int \overline{L}_{el}\pi^a dS_a L_{el}\,,\\
\label{symm1_int}\int \overline{L}_{el}{\pi'}^a  dS_a L_{el}
& = &
\sinh\theta\,\int \overline{L}_{el}\gamma^a  dS_a L_{el}
+ \cosh\theta\,\int \overline{L}_{el}\pi^a  dS_a L_{el}\,.
\end{eqnarray}
The following notations are used: ${\gamma'}^a = q^{ab}{\gamma'}_b$, 
$\gamma^a = q^{ab}\gamma_b$,
${\pi'}^a = q^{ab}{\pi'}_b$, and ${\pi}^a = q^{ab}{\pi}_b$.
One integrates over an arbitrary hypersurface with $dS_a = $
$\sqrt{-q}\,e_{abcd}\,dx^b\,d{x'}^c\,d{x''}^d$.
One may denote energy, $E = \int \overline{L}_{el}\ p^0 dV L_{el} $,
and electrical charge, $Q = \int \overline{e}_L\overline\gamma^0 dV e_L$,
where $dV \equiv dS_0$ is the element of spatial volume, and
$p^a = {\hbar\over\ell}\,\pi^a$ is the momentum operator.
Selecting spatial volume as a hypersurface of integration,
one obtains
\begin{eqnarray}\label{E}
E' &=& \cosh\theta\, E + {\hbar\over\ell}\,\sinh\theta\, Q\,;\\
\label{Q}
Q' &=& \cosh\theta\, Q + {\ell\over\hbar}\, \sinh\theta\,E\,.
\end{eqnarray}
Equations (\ref{E}) and (\ref{Q}) represent transformations
of energy and electrical charge, respectively. Suppose, that
$E = 0$. Then, $\tanh\theta = {\ell\over\hbar}\,{{E'}\over{Q'}}$,
which means
that energy $E'$ is due as a whole to electrical charge $Q'$,
placed in constant electroweak potential, as in (\ref{delta_e1}).
Thus,
\begin{equation}\label{e_L}
\tanh \theta \,=\, {\ell\over 2}\left(g'W_0^3(\infty)
+ g''B_0(\infty)\right)\,.
\end{equation}
Analogous considerations for $e_R$  lead to the equation
\begin{equation}\label{e_R}
\tanh \theta \,=\, \ell g'' B_0(\infty)\,,
\end{equation}
{}from which it follows that one should put $g'W_0^3(\infty)
= g''B_0(\infty)$.\footnote{{ }One should use
$\sin\theta_W\,g' = \cos\theta_W\,g'' = {e\over{\hbar c}}$.}
Thus, one should associate the $\theta$-transformation with that
of electroweak potentials at spatial infinity.
{}From (\ref{e_L}) (or (\ref{e_R})) it follows that $\theta \propto \ell$,
and neglecting terms of order $\ell^2$ in  (\ref{E}) and (\ref{Q}), one obtains  
$E' \approx E + \hbar g'' B_0(\infty)\, Q$, and $Q' \approx Q$.

\end{document}